\documentclass[prl,twocolumn,superscriptaddress,nofootinbib]{revtex4-1}
\usepackage{epsfig}
\usepackage{graphicx}
\usepackage{hyphenat}
\usepackage[utf8]{inputenc}
\usepackage[T1]{fontenc}
\usepackage{amsmath}
\usepackage{amssymb}
\usepackage{mathtools}
\usepackage{mathrsfs}
\usepackage{slashed}
\usepackage{epstopdf}
\usepackage{xcolor}
\usepackage{booktabs}
\usepackage{subfigure}
\graphicspath{ {image/} }
\usepackage{float}
\definecolor{lcolor}{rgb}{0.,0.0,0.}
\definecolor{citcolor}{rgb}{0,0.,0.5}
\usepackage[breaklinks,colorlinks,urlcolor=blue,citecolor=blue,linkcolor=blue]{hyperref}
\usepackage{multirow}
\usepackage{ltablex}
\usepackage{soul}
\usepackage{braket}

\newcommand{\beq}{\begin{eqnarray}}
\newcommand{\eeq}{\end{eqnarray}}

\newcommand{\bem}{\begin{multline}}
\newcommand{\eem}{\end{multline}}
\newcommand{\beg}{\begin{gather}}
\newcommand{\g}{\boldsymbol{g}}
\newcommand{\eeg}{\end{gather}}

\newcommand{\nn}{\nonumber\\}

\newcommand{\ben}{\begin{eqnarray*}}
\newcommand{\een}{\end{eqnarray*}}
\newcommand{\un}{{\boldsymbol{u}_{c}}}

\newcommand{\tvec}{\boldsymbol}
\newcommand{\na}{\nabla}

\newcommand{\pa}{\partial}

\def\0{{\mathbf{0}}}

\def\cG{{\cal G}}
\def\cP{{\cal P}}
\def\cK{{\cal K}}

\def\cC{{\cal C}}

\newcommand{\secn}[1]{Section~1}
\newcommand{\appn}[1]{Appendix~1}

\long\def\comment#1{ }

\def\and{\quad\text{and}\quad}

\def\rn{{\boldsymbol r}}

\def\q{{\boldsymbol q}}
\def\0{{\boldsymbol 0}}
\def\p{{\boldsymbol p}}
\def\l{{\boldsymbol l}}
\def\k{{\boldsymbol k}}
\def\x{{\boldsymbol x}}
\def\y{{\boldsymbol y}}
\def\X{{\boldsymbol X}}
\def\Y{{\boldsymbol Y}}

\def\u{{\boldsymbol u}}
\def\w{{\boldsymbol w}}

\newcommand{\qv}{\q}
\newcommand{\pv}{\p}
\newcommand{\lv}{\l}
\newcommand{\kv}{\k}
\newcommand{\qvb}{\bar \q}

\newcommand{\lvb}{\bar \l}
\newcommand{\kvb}{\bar \k}

\renewcommand\d{\delta}
\renewcommand\r{\rho}

\renewcommand\t{\tau}
\newcommand\m{\mu}

\newcommand{\be}{\begin{equation}}
\newcommand{\ee}{\end{equation}}
\newcommand{\bes}{\begin{subequations}}
\newcommand{\ees}{\end{subequations}}
\newcommand{\bea}{\begin{eqnarray}}
\newcommand{\eea}{\end{eqnarray}}

\newcommand\G{\mathcal{G}}

\begin{document}

\title{Quantum partonic transport in QCD matter}

\author{Jo\~ao Barata}
\email[]{jlourenco@bnl.gov}
\affiliation{Physics Department, Brookhaven National Laboratory, Upton, NY 11973, USA}
\author{Andrey V. Sadofyev}
\email[]{andrey.sadofyev@usc.es}
\affiliation{Instituto Galego de F{\'{i}}sica de Altas Enerx{\'{i}}as,  Universidade de Santiago de Compostela,\\ Santiago de Compostela 15782,  Spain}
\affiliation{NRC Kurchatov Institute, Moscow, Russia}
\author{Xin-Nian Wang}
\email[]{xnwang@lbl.gov}
\affiliation{Nuclear Science Division, MS 70R0319,
Lawrence Berkeley National Laboratory,\\ Berkeley, California 94720, USA}

\begin{abstract}
We study gradient corrections to the transport equation for energetic light partons in dense QCD environments. In the diffusion limit, the transport dynamics is solely controlled by small-angle elastic scatterings, leading to transverse momentum broadening with respect to the parton's initial direction. Such a parton propagation is usually considered in the limit of transversely homogeneous matter. The transport processes admit a classical description and the transverse spatial dependence of the medium properties emerges only through the jet quenching parameter. In this work, we show that a gradient expansion of the all-order evolution equation for the partonic Wigner function leads
to an evolution equation in the Boltzmann-diffusion form only up to the leading order in transverse gradients. At the second order in gradients, the quantum corrections associated with non-local interactions give rise to a novel transport that can be implemented in Monte Carlo simulations. In addition,  using our results, we compute the gradient corrections to the jet quenching parameter in inhomogeneous matter.

\end{abstract}

\maketitle

%%%%%%%%%%%%%%%%%%

\noindent{\bf Introduction: }Jets are commonly found in the final state of heavy-ion collisions (HIC), providing a powerful tomographic tool to study the evolution of the quark-gluon plasma (QGP) \cite{Vitev:2002pf, Apolinario:2017sob, Li:2020rqj, Arratia:2020nxw, He:2020iow, Sadofyev:2021ohn,Du:2021pqa, Antiporda:2021hpk,Barata:2022krd,Fu:2022idl, Andres:2022ndd, Sadofyev:2022hhw}. They evolve in parallel with the hot matter and are able to probe it at different time and length scales. Extracting the details of how jet structure is modified by the matter is a highly non-trivial open problem, both from the theoretical and experimental points of view~\cite{Mehtar-Tani:2013pia, Sievert:2018imd,Apolinario:2022vzg}.

For energetic colored particles, the interactions are dominated by t-channel gluon exchanges. These result in deflection of partons 
and consequent inelastic emission of gluon radiation. Describing these processes in perturbative QCD, one has to rely on multiple simplifying assumptions. For instance, treating the parton energy as the largest scale and assuming the matter to be uniform in the transverse directions, one can usually make the problem theoretically tractable in a so-called eikonal expansion, an expansion in inverse powers of energy. However, in doing so one further decouples the dynamics of the jets from the medium evolution, severely diminishing their tomographic capabilities \cite{Sadofyev:2021ohn,Barata:2022krd,Andres:2022ndd}.

Only recently, it was shown that the hydrodynamic evolution of the medium can be systematically accounted for in the parton-matter interactions \cite{Sadofyev:2021ohn}, if one goes beyond the leading eikonal limit and treats the transverse variations of the matter properties in a gradient expansion.\footnote{For earlier works introducing the hydrodynamic gradient expansion in the probe-matter interactions, see e.g. \cite{Lekaveckas:2013lha, Rajagopal:2015roa, Sadofyev:2015hxa, Li:2016bbh, Reiten:2019fta, Arefeva:2020jvo}.} Phenomenologically, the coupling of the parton evolution to the medium anisotropies reflects itself in the generation of odd moments of the underlying transverse momentum distribution, related to the non-trivial azimuthal particle distribution. So far, only the leading order gradient effects at the level of the final particle distribution have been computed~\cite{Sadofyev:2021ohn,Barata:2022krd}. In parallel, it was noticed that the same effects could be explored via the in-medium parton Wigner function \cite{Fu:2022idl}. One advantage of the Wigner function approach, in contrast with the direct computation of the particle distributions, is that it offers a direct connection to the parton evolution equation.

In the case of homogeneous matter and in the limit of small-angle in-medium scatterings, the corresponding Wigner function is a solution of the 
Boltzmann-diffusion
equation \cite{Arnold:2002ja,Arnold:2002zm,Ghiglieri:2015ala,He:2015pra}. 
Since this regime admits a classical description, one may expect that accounting for general anisotropic effects would require a full understanding of the quantum evolution of energetic partons in the medium.
One has to derive the associated quantum evolution equation and compute the corresponding Wigner function beyond the classical limit.

In this work, we derive the parton Wigner function, using the results for the parton broadening in inhomogeneous matter obtained within the BDMPS-Z approach \cite{Barata:2022krd}. We will generalize the transport equation to all orders in matter gradients, and show how it gets modified beyond the diffusion approximation. Truncating the gradient expansion up to the second order, we will show that matter anisotropies are responsible for 
novel quantum corrections to the transport equation. The resulting equation can be integrated into commonly used transport 
models for jet quenching phenomenology~\cite{Kutak:2018dim,Schlichting:2020lef,He:2015pra,JETSCAPE:2017eso,Schenke:2009gb,Blaizot:2013vha}, considerably enhancing their 
tomographic capabilities.

%%%%%%%%%%%%%%%%%%

\noindent{\bf In-medium parton Wigner function: }
In parton energy loss calculations, the QCD medium is often described with a background color field induced by stochastic color densities $\hat \r^a(\x,z)$ \cite{Gelis:2010nm,Andres:2022ndd}. Then, the leading interaction of a parton with the matter, in the limit 
when no gluon radiation is produced, can be described via a reduced single-particle propagator,
\begin{align}\label{eq:G_prop}
 \G(\x_L,L;\x_0,0)&= \int\limits_{\x_0}^{\x_L} \mathcal D\rn  \exp\left(\frac{iE}{2}\int\limits_{0}^{L}  d\t \, \dot{\rn}^2\right)\notag\\
 &\hspace{0cm}\times\cP\exp\left(i\int\limits_{0}^{L}  d\t \, t^a_{\text{proj}}v^a(\rn(\t),\t)\right) \,,  
\end{align}
where we have assumed that the sources are in the same color representation $R$ \cite{Barata:2022krd, Andres:2022ndd},
$E$ is the conserved energy of the parton, $L$ is the longitudinal size of the matter, and $\x_L$ and $\x_0$ are the initial and final
two dimensional coordinates in the directions transverse to the initial parton's momentum. 
In this picture, the parton-matter interactions are described by an effective potential $v^a(\x, z)=\int_\q \,e^{i\q\cdot\x}v(\q^2, z)\,\r^a(\q, z)$.\footnote{Throughout this work, we will use the shorthand notations $\int d^2\q/(2\pi)^2=\int_\q$ and $\int d^2\x=\int_\x$.} 
In turn, the 
elementary scattering potential $v(\q^2, z)$ corresponds to the particular matter model, defining how the in-medium gluon field is screened at large distances, and its $z$-dependence is driven by the screening mass. The screening length is assumed to be small comparing to the characteristic distance between the sources.

The presence of the matter results in the broadening of the transverse momentum distribution of the partons. This process can be described through
the corresponding parton Wigner function~\cite{Fu:2022idl,He:2020iow}, which reads
\begin{align}
\label{eq:Wigner}
 &W_L(\Y,\p)\equiv \int_{\y,\x,\X,\p_0}\,e^{-i\left(\p\cdot\y-\p_0\cdot\x\right)}\,W_0(\X,\p_0)\notag\\
 &\hspace{0.5cm}\times\left\langle \cG\left(\Y+\frac{\y}{2};\X+\frac{\x}{2}\right)\cG^\dagger\left(\Y-\frac{\y}{2};\X-\frac{\x}{2}\right)\right\rangle\,.
\end{align}
Here we have used shorthand notations omitting the length dependence in the propagators,
$\Y$ can be interpreted as an impact parameter of an effective dipole formed by the parton in the amplitude and its complex conjugate, and $\p$ is the parton's final transverse momentum. 
Also, we implicitly assume that the medium averaging involves a trace over the color indices, such that $\langle{\bf I}\rangle=1$ with ${\bf I}$ being the identity matrix in the color space. The initial Wigner function 
$W_0(\X,\p_0)$, is naturally related to the initial distribution of partons. Notice that the Wigner function should include a gauge link. However, here this link is trivial since the transverse field components are zero for static sources in the Lorentz gauge~\cite{Fu:2022idl}.

The two-point correlator of the propagators entering \eqref{eq:Wigner} implies an averaging over the scattering centers. In this work and as is usually done, we assume that the medium can be described in terms of classical background densities with Gaussian statistics~\cite{Blaizot:2015lma,Mehtar-Tani:2013pia,Qin:2015srf}
\be
\left\langle\hat\rho^a(\q,z)\hat\rho^b(\bar \q,\bar z)\right\rangle 
= \frac{\delta^{ab}}{2C_{\bar R}}\delta(z-\bar z)
\,\rho(\q+\bar \q,z)\,,
\label{eq:FTaverage}
\ee
where $\rho(\q+\bar \q,z)$ is a mixed representation of $\rho(\x, z)$, the number density of the sources, and $C_{\bar{R}}$ is the quadratic Casimir of the representation opposite to the representation of the sources. 
We further follow \cite{Barata:2022krd,Sadofyev:2021ohn,Andres:2022ndd}, assuming that the medium properties are slowly varying from point to point. 
This allows one to study $W$ using an expansion in the transverse gradients of the medium parameters, such as the density of the scattering centers $\rho(\tvec{x})$ and the Debye mass $\mu(\tvec{x})$ in the scattering potential. To do so, we will Taylor expand the parameters
\begin{align}\label{eq:grad_exp}
    &\rho(\x) \approx \rho + \x\cdot \tvec\na\rho\, +\frac{1}{2}x^i x^j \na_i\na_j\,\r\, , \nn
    &\mu^2 (\x) \approx \mu^2 + \x\cdot\tvec\na\mu^2 \, +\frac{1}{2}x^i x^j \na_i\na_j\,\m^2 \,, 
\end{align}
assuming for simplicity that they (and their transverse gradients) are $z$-independent.

At the leading order in gradients, the two-point function of the single-particle propagators has been previously computed in \cite{Barata:2022krd}, and reads
\begin{align}
\label{eq:GGfinal}
 &\left\langle \cG(\x_L;\x_0)\cG^\dagger(\overline{\x}_L;\overline{\x}_0)\right\rangle
 =\left(\frac{E}{2\pi L}\right)^2\notag\\
 &\hspace{1cm}\times\frac{\exp\left\{iE\left( \tvec{w}\cdot\dot{\u}_c\right)\Big|_0^L-\int\limits_0^L\,d\t\,\mathcal{V}\left(\un(\t)\right)\right\}}{1+\frac{i}{EL}\hat{\g}\cdot\int\limits^L_0d\zeta\int\limits^\zeta_0d\xi\,\xi\,\tvec{\na}\mathcal{V}\left(\un(\xi)\right)} \, ,
\end{align}
where $\w\equiv\frac{\x+\bar{\x}}{2}$, $\u\equiv\x-\bar{\x}$, $\mathcal{V}(\q)$ is the 
dipole potential, and $\hat \g \equiv \tvec{\na}\rho \frac{\delta }{\delta \rho} +\tvec{\na }\mu^2 \frac{\delta }{\delta \mu^2}$ is an operator generating the gradient corrections. The trajectory $\u_c(\t)$ is the solution of the classical equations of motion, arising in the path integral.\footnote{To leading order in transverse gradients, see ~\cite{Barata:2022krd}, 
$\u_c=\u_c^{(0)}+\u_c^{(1)}$ with $\u_c^{(0)}(\t)=\t \, (\boldsymbol{u}_L-\boldsymbol{u}_0)/L +\boldsymbol{u}_0$ and 
\begin{align*}
\u_c^{(1)}(\t)&=\frac{i}{E}\,\hat{\tvec{g}}\Bigg\{\int\limits^\t_0\,d\zeta\,\int\limits_0^\zeta\,d\xi -\frac{\t}{L}\int\limits^L_0\,d\zeta\,\int\limits_0^\zeta\,d\xi\Bigg\} \mathcal{V}\left(\u_c^{(0)}(\xi)\right)\,. 
\end{align*}
} The dipole potential is defined as 
$$
\mathcal V(\tvec{q})\equiv -C\,\r\left(\left|v(\q^2)\right|^2-(2\pi)^2\delta^{(2)}(\tvec{q})\int_{\tvec{l}}\,\left|v(\l^2)\right|^2\right)\,,
$$
where $C=\frac{C_{proj}}{2C_{\bar{R}}}$
, and $C_{proj}$ is the quadratic Casimir in the representation of the energetic parton.

We are interested in the diffusion regime, which corresponds to the so-called harmonic approximation for $\mathcal{V}(\q)$. In this regime, the soft gluon exchanges with the medium are captured but the hard momentum transfers are
neglected, and the dipole potential can be approximated by
\begin{align}\label{eq:V_harm}
\mathcal{V}(\y)\equiv \int_\q e^{i\q \cdot \y} \,  \mathcal{V}(\q)\approx \frac{\hat q}{4}    \y^2\, ,
\end{align}
where $\hat{q}\equiv\hat{q}_0 \log \frac{Q^2}{\mu^2}$ is the jet quenching parameter, and $Q$ is a free large momentum scale ubiquitous to the harmonic approximation. The explicit form of $\hat{q}_0$ in terms of medium parameters depends on the model of the medium. Here we will use the Gyulassy-Wang (GW) model, corresponding to $v(\q^2)=-g^2/(\q^2+\m^2)$
with $g$ the strong coupling constant and the \textit{bare} jet quenching coefficient $\hat{q}_0  \equiv 4\pi\,C\, \alpha_s^2  \rho$. 
For other medium models such as the ones based on the Hard Thermal Loops approximation~\cite{Aurenche:2002pd} or in holographic pictures~\cite{Liu:2006he}, Eq.~\eqref{eq:V_harm} still holds, though the dependence of $\hat q_0$ on the medium parameters is different~\cite{Barata:2020rdn}.

Under all these assumptions, the resulting Wigner function can be easily obtained using \eqref{eq:GGfinal}. For point-like initial conditions, $W_0(\X,\p_0)=(2\pi)^2\d^{(2)}(\X)\d^{(2)}(\p_0)$, the final state Wigner function can be written as
\begin{widetext}
\begin{align}\label{eq:Wigner_function}
 &W(\Y,\p)=\frac{48 E^2}{\hat q^2L^4}\exp\left\{-\frac{4}{\hat qL^3}\left(L^2\p^2-3E L\,\p \cdot \Y+3 E^2\Y^2\right)\right\}\notag\\
 &\hspace{2.5cm}\times\Bigg[1+\frac{\g}{15E\hat{q}L^3}\cdot\Big(\hat qL^4 \,\p-5L^3\left(\p^2\,\p+3E\hat{q}\,\Y\right)+12E L^2\left(4\p^2\,\Y+(\p\cdot\Y)\p\right) \nn & \hspace{5cm}
 +90E^3\Y^2\,\Y-9E^2L\left(14(\p\cdot\Y)\Y+\Y^2\,\p\right)\Big)\Bigg]\,,
\end{align}
\end{widetext}
where $\g$ is the parameter resulting from $\hat \g$ which encapsulates the matter gradients. For the GW model in the harmonic approximation, the gradient parameter can be written as 
\begin{align}
 \g\equiv\frac{\tvec{\na}\hat{q}}{\hat{q}}=  \frac{\tvec{\na}\rho}{\rho}-\frac{1}{\log \frac{Q^2}{\mu^2}}\frac{\tvec{\na}\mu^2}{\mu^2}\, .
\end{align}

%%%%%%%%%%%%%%%%%%

\noindent{\bf Kinetic approach to parton transport: } 
While we have derived the final state Wigner distribution to the leading order in gradients \eqref{eq:Wigner_function}, for many phenomenological applications the evolution equation defining $W$ is of greater importance, since it allows to study the in-medium parton evolution within the kinetic picture, which is more suitable for numerical simulations.
In the past, it has been noticed that in the absence of gradients and in the small angle scattering approximation the Wigner function satisfies a diffusion equation \cite{He:2020iow}. In the present case, having access to the explicit form of the Wigner function \eqref{eq:Wigner_function}, we can directly check how the leading order gradients modify the diffusion in an inhomogeneous matter.

In the kinetic approach to the in-medium evolution of the parton
distribution $f(\Y,\p)$, one usually starts from the following reduced Boltzmann equation~\cite{Arnold:2002ja,Arnold:2002zm}
\begin{align}
\left(\partial_L + \frac{\p}{E}\cdot \tvec{\na}_\Y\right) f(\Y,\p) = \cC[f] \, ,
\end{align}
where $\cC[f]$ is the collision kernel, involving only energy conserving elastic processes in our approximation, and the force terms are assumed to be  vanishing. 
To generalize the result to the inhomogeneous case, 
one has to take into account the impact parameter dependence of the parton-matter interactions within a detailed \textit{microscopic} derivation.

It is instructive to notice the following two facts. First, one may attempt to generalize the evolution equation in a naive ideal-hydrodynamic way, promoting the only kinetic coefficient $\hat{q}$ to a function of $\Y$. Second, we already have the solution for the Wigner function, derived with the field theoretical methods. Thus, we can substitute 
\eqref{eq:Wigner_function} into the naively generalized evolution equation, and, expanding $\hat q (\Y)\simeq \hat q + \Y \cdot \tvec{\na}\hat{q}$, we find 
\begin{align}\label{eq:final_kinetic}
&\left(\partial_L + \frac{\p}{E}\cdot \tvec{\na}_\Y-\frac{\hat{q}(\Y)}{4}\partial^2_\p\right) W(\Y,\p) 
= \mathcal{O}\left(\pa^2_\perp\hat{q}\right) \, .
\end{align}

As expected, in the absence of the gradient corrections,
the Wigner functions obeys the usual Boltzmann-diffusion equation, with $\hat q$ playing the role of the diffusion constant. Remarkably, the leading order gradient corrections in \eqref{eq:Wigner_function} are accounted for by 
the minimal replacement $\hat q \to \hat q (\Y)$, 
and the naive generalization of the evolution equation does work at this order. This observation indicates that the kinetic description of momentum broadening gains no structural corrections at the first order in gradients, and the two approaches to the problem are equivalent at this level.

However, the naive generalization of 
the evolution equation is not necessarily adequate in general. Indeed, assuming that the in-medium scattering rates are local in impact parameter 
implies that the resulting transport equation has the same functional form as the usual Boltzmann-diffusion equation (up to a drift term), with the minimal replacement $\hat q \to \hat q(\Y)$. 
This locality assumption can be lifted 
if one considers the 
quantum generalization of Boltzmann transport, e.g. starting from  
the Kadanoff-Baym 
equations~\cite{PhysRev.124.287}.
In turn, the results in \cite{Barata:2022krd} indicate that the higher order gradient corrections to \eqref{eq:GGfinal} are non-trivial, and the two approaches should be further compared. The fact that the given Wigner function \eqref{eq:Wigner_function} solves the particular evolution equation is only indicative.

%%%%%%%%%%%%%%%%%%

\noindent{\bf Quantum corrections to parton transport:} In the previous section, we have discussed how the leading order gradient correction to the evolution equation can be absorbed into a spatial dependent jet quenching parameter. 
Now, we turn to the derivation of the evolution equation beyond the first order in gradients from a microscopic consideration.
For simplicity, we will focus on the gradients of $\rho$, while $\mu$ gradients can be obtained similarly.

Since we are interested in the evolution equation governing the Wigner function, it is instructive to study the path length derivative of the two-point correlator in \eqref{eq:GGfinal}.
For that, we consider the two-point function of the propagators between $z=0$ and $z=L+\epsilon$ with 
$\epsilon\ll L$. 
Following \cite{Andres:2022ndd}, one can express it 
as a convolution
\begin{align}
\label{eq:GG}
&\left\langle \mathcal{G}^\dagger(\bar \k,L+\epsilon; \bar \k_0, 0)\,
\mathcal{G}(\k,L+\epsilon;\k_0,0) \right\rangle \nn 
&\hspace{1cm}= \int_{\l,\bar \l}
\left\langle \mathcal{G}^\dagger(\bar \k,L+\epsilon; \bar \l,L)\,\mathcal{G}(\k,L+\epsilon;\l,L)\right\rangle \nn
&\hspace{2cm}\times \left\langle \mathcal{G}^\dagger(\bar \l,L;\bar \k_0,0)\,\mathcal{G}(\l,L;\k_0,0)\right\rangle\,, 
\end{align}
where we use the locality of the averages in $z$ to break the full evolution into two steps, both in color 
singlet states.

Expanding \eqref{eq:GG} to the leading order in $\epsilon$, one can extract the evolution kernel. We then find that the Wigner function in a momentum space representation satisfies
\begin{align}
\label{eq:DiffEqGeneral}
\pa_L W(\k,\bar \k)&=-i\,\frac{\k^2-\bar \k^2}{2E}W(\k,\bar \k)\nn 
&\hspace{1cm}-\int_{\q,\bar \q, \l,\bar \l} \,\cK(\q,\bar \q;\l,\bar \l )W(\l,\bar \l)\,,
\end{align}
where 
\begin{align}
&\cK(\q,\bar \q;\lv,\lvb)= -(2\pi)^4\, C\, v(\qv)v(\qvb)\nn
&\hspace{0.15cm}\times\Bigg\{\rho(\qv-\qvb)\,\delta^{(2)}(\kv-\qv-\lv)\delta^{(2)}(\kvb-\qvb-\lvb)\notag\\
&\hspace{0.3cm}-\frac{1}{2}\rho(\qv+\qvb)\,\delta^{(2)}(\kv-\lv)\delta^{(2)}(\kvb-\qv-\qvb-\lvb)\notag\\
&\hspace{0.45cm}-\frac{1}{2}\rho^\dag(\qv+\qvb)\,\delta^{(2)}(\kv-\qv-\qvb-\lv)\delta^{(2)}(\kvb-\lvb)\Bigg\}\,,
\end{align}
and $\rho(\k)$ is the Fourier of the source number density. 

The evolution equation \eqref{eq:DiffEqGeneral} generalizes \eqref{eq:final_kinetic} for exact transverse coordinate dependence of the matter parameters, and is valid beyond the harmonic approximation. Expanding \eqref{eq:DiffEqGeneral} in gradients, we substitute the explicit form of 
\eqref{eq:grad_exp} up to the second order.
Using the shorthand notations
$h'(\q^2)\equiv \partial_{\q^2} h(\q^2)$ and  $\p=(\k + \bar \k)/2$, we write the second order evolution equation in the harmonic approximation as
% %
\begin{align}\label{eq:main_result}
&\left(\pa_L+\frac{\pv\cdot\tvec{\na}_\Y}{E}-\frac{\hat{q}(\tvec{Y})}{4}\pa_\p^2\right) W(\Y,\pv)=\na_i\na_j\rho\nn
&\hspace{0.5cm}\times\int_{\q}\left[\kappa \frac{\pa^2}{\pa p_i\pa p_j} \d^{(2)}(\qv)-V_{ij}(\q)\right]W(\Y,\pv-\qv)\, ,
\end{align}
% %
where we have introduced $\kappa    = 2\pi^2C\int_{\q} v^2$ and a new directional potential reading
\begin{align}
&V_{ij}(\qv)=\frac{C}{2}\Bigg(\Big\{2q_i q_j \left[vv^{\prime\prime}-v^\prime v^\prime\right]+vv^\prime\delta_{ij}\Big\}\notag\\
&\hspace{0.25cm}-(2\pi)^2\d^{(2)}(\qv)\int_\lv\Big\{2l_i l_j \left[v v^{\prime\prime}-v^\prime v^\prime\right]+vv^\prime\delta_{ij}\Big\}\Bigg)\,.
\end{align}
The term proportional to $\rho (\Y)$ in \eqref{eq:main_result} is a direct generalization of its analogue in \eqref{eq:final_kinetic}. However, at the second order the evolution equation gains additional terms of new functional form. 

Notice that these do not directly depend on the effective scattering potential $\mathcal{V}$, but rather on the in-medium scattering cross-section $v$. As a consequence, Eq.~\eqref{eq:main_result}, even in the diffusion approximation, is more sensitive to the details of the medium model.

Another important conclusion is that assuming locality of interactions in $\Y$, one implies that all corrections to the evolution equation could be taken into account with the minimal replacement $\hat q\to \hat q(\Y)$. Thus, one may conclude that the additional terms in \eqref{eq:main_result} 
can only be obtained if the interactions between the matter 
and the jet 
are non-local in $\Y$. 
These types of interactions are linked to the quantum nature of the evolution of the parton in the medium, and thus the novel evolution equation also accounts for quantum corrections to the classical Boltzmann transport, see e.g. \cite{Weickgenannt:2021cuo,Sheng:2021kfc} for a recent discussion.

Finally, it is instructive to look at the second moment of the Wigner function, since it provides access  to the full jet quenching parameter $\hat q_r\equiv \partial_L \langle  \p^2 \rangle$ containing power corrections induced by the matter gradients. 
The terms in \eqref{eq:main_result} depending on the scattering rates can be absorbed into a coefficient $\eta = \rho \,  \kappa/(2\pi^2\hat q) + \frac{C \rho}{2\hat{q}} \int_\q \q^2
v^2 [\q^2v'/v ]'$, 
which can be computed directly once 
a particular model for the medium is chosen. Using 
that at the zeroth order in gradients 
$ \int_{\Y,\p} Y_i Y_j W(\Y,\p)= \delta_{ij} \, \hat q  L^3/(6 E^2)$, we find the full jet quenching parameter
\begin{align}\label{eq:qr}
 \hat q_r = \hat q  + \tvec{\na}^2 \hat q \left(\frac{\hat q L^3}{12 E^2} +  \eta \right)  \, .
\end{align}
This effective  jet quenching coefficient gets two types of corrections. The first ones are the sub-sub-eikonal terms, which one would expect since the gradient expansion is sensitive to the kinetic phases~\cite{Barata:2022krd,Sadofyev:2021ohn}. 
Such corrections will compete with other sub-sub-eikonal effects we did not include, although with different parametric dependence. 
In turn, the $\eta$-term on the right-hand side of \eqref{eq:qr} is independent of 
$E$, and enters at the eikonal accuracy. 
Similar energy independent corrections have recently been observed in non-perturbative real-time simulations of single parton in-medium evolution~\cite{Barata:2022wim}.
The phenomenological implications of these novel corrections require further study, which we leave for future work.

%%%%%%%%%%%%%%%%%%

\noindent{\bf Conclusion:} In this letter, we have detailed the first \textit{ab-initio} derivation of the kinetic equation describing the evolution of energetic partons in inhomogeneous QCD environments. Studying the gradient corrections to the in-medium parton Wigner function, we have derived the 
all-order master transport equation. Although the leading gradient corrections can be accounted for in the Boltzmann-diffusion form 
under a minimal replacement for $\hat q$, already at the second order 
the resulting equation has functionally novel terms. 
We have further argued 
that these are associated with non-local interactions, and 
can be considered as quantum corrections to the classical transport. As a direct consequence of 
the modified evolution, 
we find new power corrections to the jet quenching parameter. 
Strikingly, 
$\hat{q}$ gets energy independent contributions, which could be sizeable compared to radiative corrections~\cite{Caucal:2022fhc,Blaizot:2014bha,Liou:2013qya,Ghiglieri:2022gyv}.

The new transport equation derived in this paper has ample applications in jet quenching phenomenology, allowing to include gradient effects in existing parton transport models.
Besides such phenomenological applications,  
it would be important to confirm that using 
the formal kinetic theory approach, it is possible to recover the master equation derived in this paper. Such an exercise would give new insights into the role played by matter gradients, helping to lift some of the  
assumptions. 
Another interesting extension would be to include other sub-eikonal corrections, such as friction terms, into the evolution equation. These corrections would compete with the gradient terms and are needed to have a complete picture of in-medium evolution. 

The results derived here are applicable in the QGP phase of HIC, but the formalism can be extended to the other phases of QCD matter. Since anisotropic effects might be more important at earlier times in the aftermath of HIC or in smaller collision systems~\cite{Hauksson:2021okc,Ipp:2020mjc,Carrington:2021dvw,Brewer:2021kiv}, having theoretical tools to describe such scenarios is of utmost importance. The inclusion of inelastic processes can also be studied using the approach followed in this work, although its exact kinetic formulation is not known at the moment for inhomogeneous backgrounds.

%%%%%%%%%%%%%%%%%%
\noindent{\bf Acknowledgments:} %JB is grateful to 
The authors would like to thank J.-P. Blaizot, X. Mayo, Y. Mehtar-Tani, C. Salgado and E. Speranza for helpful discussions. JB was supported by DOE under Contract No.~DE-SC0012704. The work of AVS was funded by the Russian Science Foundation Grant 22-22-00664. AVS is also grateful for support from European Research Council project ERC-2018-ADG-835105 YoctoLHC; from Maria de Maetzu excellence program under project MDM-2016-0692 and CEX2020-001035-M; from Spanish Research State Agency under project PID2020-119632GB-I00; from Xunta de Galicia (Centro singular de investigación de Galicia accreditation 2019-2022), and from European Union ERDF. XNW was supported by DOE under  Contract No. DE-AC02-05CH11231, by NSF under Grant No. ACI-1550228 within the JETSCAPE and OAC-2004571 within the X-SCAPE Collaboration.
The authors are grateful to 
the workshop "Jet Quenching in the Quark Gluon Plasma", where this work was initiated. This work has been also supported by STRONG-2020 "The strong interaction at the frontier of knowledge: fundamental research and applications" which received funding from the European Union’s Horizon 2020 research and innovation program under grant agreement No 824093.

%%%%%%%%%%%%%%%%%%%%%%%%%%%%%%%%%%%%%%%%%%%%%%%%%%%%%%%%%%%%%%%%%%%%%%%%%%%%%%%%%%%%%%%%%%
%%%%%%%%%%%%%%%%%%%%%%%%%%%%%%%%%%%%%%%%%%%%%%%%%%%%%%%%%%%%%%%%%%%%%%%%%%%%%%%%%%%%%%%%%%

\bibliographystyle{apsrev4-1}
\bibliography{references.bib}

\end{document}